
\input phyzzx.tex
\catcode`\@=11
\def\wlog#1{}
\def\eqname#1{\rel@x {\pr@tect
  \ifnum\equanumber<0 \xdef#1{{\rm\number-\equanumber}}%
     \gl@bal\advance\equanumber by -1
  \else \gl@bal\advance\equanumber by 1
     \ifx\chapterlabel\rel@x \def\d@t{}\else \def\d@t{.}\fi
    \xdef#1{{\rm\chapterlabel\d@t\number\equanumber}}\fi #1}}
\catcode`\@=12
\catcode`\@=11

\def\eat@#1{}
\mathchardef\prime@="0230
\def\prime{{{}\prime@{}}}
\def\prim@s{\prime@\futurelet\next\pr@m@s}

\def\,{\relax\ifmmode\mskip\thinmuskip\else\thinspace\fi}
\def\!{\relax\ifmmode\mskip-\thinmuskip\else\negthinspace\fi}
\def\frac#1#2{{#1\over#2}}
\def\dfrac#1#2{{\displaystyle{#1\over#2}}}

\def\:{\nobreak\hskip.1111em{:}\hskip.3333em plus .0555em\relax}
\def\intic@{\mathchoice{\hskip5\p@}{\hskip4\p@}{\hskip4\p@}{\hskip4\p@}}
\def\negintic@
 {\mathchoice{\hskip-5\p@}{\hskip-4\p@}{\hskip-4\p@}{\hskip-4\p@}}
\def\intkern@{\mathchoice{\!\!\!}{\!\!}{\!\!}{\!\!}}
\def\intdots@{\mathchoice{\cdots}{{\cdotp}\mkern1.5mu
    {\cdotp}\mkern1.5mu{\cdotp}}{{\cdotp}\mkern1mu{\cdotp}\mkern1mu
      {\cdotp}}{{\cdotp}\mkern1mu{\cdotp}\mkern1mu{\cdotp}}}
\newcount\intno@
\def\iint{\intno@=\tw@\futurelet\next\ints@}
\def\iiint{\intno@=\thr@@\futurelet\next\ints@}
\def\iiiint{\intno@=4 \futurelet\next\ints@}
\def\idotsint{\intno@=\z@\futurelet\next\ints@}
\def\ints@{\findlimits@\ints@@}
\newif\iflimtoken@
\newif\iflimits@
\def\findlimits@{\limtoken@false\limits@false\ifx\next\limits
 \limtoken@true\limits@true
   \else\ifx\next\nolimits\limtoken@true\limits@false
    \fi\fi}
\def\multintlimits@{\intop\ifnum\intno@=\z@\intdots@
  \else\intkern@\fi
    \ifnum\intno@>\tw@\intop\intkern@\fi
     \ifnum\intno@>\thr@@\intop\intkern@\fi\intop}
\def\multint@{\int\ifnum\intno@=\z@\intdots@\else\intkern@\fi
   \ifnum\intno@>\tw@\int\intkern@\fi
    \ifnum\intno@>\thr@@\int\intkern@\fi\int}
\def\ints@@{\iflimtoken@\def\ints@@@{\iflimits@
   \negintic@\mathop{\intic@\multintlimits@}\limits\else
    \multint@\nolimits\fi\eat@}\else
     \def\ints@@@{\multint@\nolimits}\fi\ints@@@}
\def\Sb{_\bgroup\vspace@
        \baselineskip=\fontdimen10 \scriptfont\tw@
        \advance\baselineskip by \fontdimen12 \scriptfont\tw@
        \lineskip=\thr@@\fontdimen8 \scriptfont\thr@@
        \lineskiplimit=\thr@@\fontdimen8 \scriptfont\thr@@
        \Let@\vbox\bgroup\halign\bgroup \hfil$\scriptstyle
            {##}$\hfil\cr}
\def\endSb{\crcr\egroup\egroup\egroup}
\def\Sp{^\bgroup\vspace@
        \baselineskip=\fontdimen10 \scriptfont\tw@
        \advance\baselineskip by \fontdimen12 \scriptfont\tw@
        \lineskip=\thr@@\fontdimen8 \scriptfont\thr@@
        \lineskiplimit=\thr@@\fontdimen8 \scriptfont\thr@@
        \Let@\vbox\bgroup\halign\bgroup \hfil$\scriptstyle
            {##}$\hfil\cr}
\def\endSp{\crcr\egroup\egroup\egroup}
\def\Let@{\relax\iffalse{\fi\let\\=\cr\iffalse}\fi}
\def\vspace@{\def\vspace##1{\noalign{\vskip##1 }}}
\def\aligned{\,\vcenter\bgroup\vspace@\Let@\openup\jot\m@th\ialign
  \bgroup \strut\hfil$\displaystyle{##}$&$\displaystyle{{}##}$\hfil\crcr}
\def\endaligned{\crcr\egroup\egroup}
\def\matrix{\,\vcenter\bgroup\Let@\vspace@
    \normalbaselines
  \m@th\ialign\bgroup\hfil$##$\hfil&&\quad\hfil$##$\hfil\crcr
    \mathstrut\crcr\noalign{\kern-\baselineskip}}
\def\endmatrix{\crcr\mathstrut\crcr\noalign{\kern-\baselineskip}\egroup
                \egroup\,}
\newtoks\hashtoks@
\hashtoks@={#}
\def\format{\crcr\egroup\iffalse{\fi\ifnum`}=0 \fi\format@}
\def\format@#1\\{\def\preamble@{#1}%
  \def\c{\hfil$\the\hashtoks@$\hfil}%
  \def\r{\hfil$\the\hashtoks@$}%
  \def\l{$\the\hashtoks@$\hfil}%
  \setbox\z@=\hbox{\xdef\Preamble@{\preamble@}}\ifnum`{=0 \fi\iffalse}\fi
   \ialign\bgroup\span\Preamble@\crcr}

\def\cases{\left\{\,\vcenter\bgroup\vspace@
     \normalbaselines\openup\jot\m@th
       \Let@\ialign\bgroup$##$\hfil&\quad$##$\hfil\crcr
      \mathstrut\crcr\noalign{\kern-\baselineskip}}

\newif\iftagsleft@
\tagsleft@true
\def\TagsOnRight{\global\tagsleft@false}
\def\tag#1$${\iftagsleft@\leqno\else\eqno\fi
 \hbox{\def\pagebreak{\global\postdisplaypenalty-\@M}%
 \def\nopagebreak{\global\postdisplaypenalty\@M}\rm(#1\unskip)}%
  $$\postdisplaypenalty\z@\ignorespaces}
\interdisplaylinepenalty=\@M
\def\allowdisplaybreak@{\def\allowdisplaybreak{\noalign{\allowbreak}}}
\def\displaybreak@{\def\displaybreak{\noalign{\break}}}
\def\align#1\endalign{\def\tag{&}\vspace@\allowdisplaybreak@\displaybreak@
  \iftagsleft@\lalign@#1\endalign\else
   \ralign@#1\endalign\fi}
\def\ralign@#1\endalign{\displ@y\Let@\tabskip\centering
   \halign to\displaywidth
     {\hfil$\displaystyle{##}$\tabskip=\z@&$\displaystyle{{}##}$\hfil
       \tabskip=\centering&\llap{\hbox{(\rm##\unskip)}}\tabskip\z@\crcr
             #1\crcr}}
\def\lalign@
 #1\endalign{\displ@y\Let@\tabskip\centering\halign to \displaywidth
   {\hfil$\displaystyle{##}$\tabskip=\z@&$\displaystyle{{}##}$\hfil
   \tabskip=\centering&\kern-\displaywidth
        \rlap{\hbox{(\rm##\unskip)}}\tabskip=\displaywidth\crcr
               #1\crcr}}
\def\overrightarrow{\mathpalette\overrightarrow@}
\def\overrightarrow@#1#2{\vbox{\ialign{$##$\cr
    #1{-}\mkern-6mu\cleaders\hbox{$#1\mkern-2mu{-}\mkern-2mu$}\hfill
     \mkern-6mu{\to}\cr
     \noalign{\kern -1\p@\nointerlineskip}
     \hfil#1#2\hfil\cr}}}
\def\overleftarrow{\mathpalette\overleftarrow@}
\def\overleftarrow@#1#2{\vbox{\ialign{$##$\cr
     #1{\leftarrow}\mkern-6mu\cleaders
      \hbox{$#1\mkern-2mu{-}\mkern-2mu$}\hfill
      \mkern-6mu{-}\cr
     \noalign{\kern -1\p@\nointerlineskip}
     \hfil#1#2\hfil\cr}}}
\def\overleftrightarrow{\mathpalette\overleftrightarrow@}
\def\overleftrightarrow@#1#2{\vbox{\ialign{$##$\cr
     #1{\leftarrow}\mkern-6mu\cleaders
       \hbox{$#1\mkern-2mu{-}\mkern-2mu$}\hfill
       \mkern-6mu{\to}\cr
    \noalign{\kern -1\p@\nointerlineskip}
      \hfil#1#2\hfil\cr}}}
\def\underrightarrow{\mathpalette\underrightarrow@}
\def\underrightarrow@#1#2{\vtop{\ialign{$##$\cr
    \hfil#1#2\hfil\cr
     \noalign{\kern -1\p@\nointerlineskip}
    #1{-}\mkern-6mu\cleaders\hbox{$#1\mkern-2mu{-}\mkern-2mu$}\hfill
     \mkern-6mu{\to}\cr}}}
\def\underleftarrow{\mathpalette\underleftarrow@}
\def\underleftarrow@#1#2{\vtop{\ialign{$##$\cr
     \hfil#1#2\hfil\cr
     \noalign{\kern -1\p@\nointerlineskip}
     #1{\leftarrow}\mkern-6mu\cleaders
      \hbox{$#1\mkern-2mu{-}\mkern-2mu$}\hfill
      \mkern-6mu{-}\cr}}}
\def\underleftrightarrow{\mathpalette\underleftrightarrow@}
\def\underleftrightarrow@#1#2{\vtop{\ialign{$##$\cr
      \hfil#1#2\hfil\cr
    \noalign{\kern -1\p@\nointerlineskip}
     #1{\leftarrow}\mkern-6mu\cleaders
       \hbox{$#1\mkern-2mu{-}\mkern-2mu$}\hfill
       \mkern-6mu{\to}\cr}}}
\def\sqrt#1{\radical"270370 {#1}}
\def\dots{\relax\ifmmode\let\next=\ldots\else\let\next=\tdots@\fi\next}
\def\tdots@{\unskip\ \tdots@@}
\def\tdots@@{\futurelet\next\tdots@@@}
\def\tdots@@@{$\mathinner{\ldotp\ldotp\ldotp}\,
   \ifx\next,$\else
   \ifx\next.\,$\else
   \ifx\next;\,$\else
   \ifx\next:\,$\else
   \ifx\next?\,$\else
   \ifx\next!\,$\else
   $ \fi\fi\fi\fi\fi\fi}
\def\text{\relax\ifmmode\let\next=\text@\else\let\next=\text@@\fi\next}
\def\text@@#1{\hbox{#1}}
\def\text@#1{\mathchoice
 {\hbox{\everymath{\displaystyle}\def\textfonti{\the\textfont1 }%
    \def\textfontii{\the\textfont2 }\textdef@@ T#1}}
 {\hbox{\everymath{\textstyle}\def\textfonti{\the\textfont1 }%
    \def\textfontii{\the\textfont2 }\textdef@@ T#1}}
 {\hbox{\everymath{\scriptstyle}\def\textfonti{\the\scriptfont1 }%
   \def\textfontii{\the\scriptfont2 }\textdef@@ S\rm#1}}
 {\hbox{\everymath{\scriptscriptstyle}%
   \def\textfonti{\the\scriptscriptfont1 }%
   \def\textfontii{\the\scriptscriptfont2 }\textdef@@ s\rm#1}}}
\def\textdef@@#1{\textdef@#1\rm \textdef@#1\bf
   \textdef@#1\sl \textdef@#1\it}

\def\textdef@#1#2{%
 \def\next{\csname\expandafter\eat@\string#2fam\endcsname}%
\if S#1\edef#2{\the\scriptfont\next\relax}%
 \else\if s#1\edef#2{\the\scriptscriptfont\next\relax}%
 \else\edef#2{\the\textfont\next\relax}\fi\fi}
\scriptfont\itfam=\tenit \scriptscriptfont\itfam=\tenit
\scriptfont\slfam=\tensl \scriptscriptfont\slfam=\tensl
\mathcode`\0="0030
\mathcode`\1="0031
\mathcode`\2="0032
\mathcode`\3="0033
\mathcode`\4="0034
\mathcode`\5="0035
\mathcode`\6="0036
\mathcode`\7="0037
\mathcode`\8="0038
\mathcode`\9="0039
\def\Cal{\relax\ifmmode\let\next=\Cal@\else
    \def\next{\errmessage{Use \string\Cal\space only in %
      math mode}}\fi\next}
    \def\Cal@#1{{\fam2 #1}}
\def\bold{\relax\ifmmode\let\next=\bold@\else
    \def\next{\errmessage{Use \string\bold\space only in %
      math mode}}\fi\next}
    \def\bold@#1{{\fam\bffam #1}}
\mathchardef\Gamma="0000
\mathchardef\Delta="0001
\mathchardef\Theta="0002
\mathchardef\Lambda="0003
\mathchardef\Xi="0004
\mathchardef\Pi="0005
\mathchardef\Sigma="0006
\mathchardef\Upsilon="0007
\mathchardef\Phi="0008
\mathchardef\Psi="0009
\mathchardef\Omega="000A
\mathchardef\varGamma="0100
\mathchardef\varDelta="0101
\mathchardef\varTheta="0102
\mathchardef\varLambda="0103
\mathchardef\varXi="0104
\mathchardef\varPi="0105
\mathchardef\varSigma="0106
\mathchardef\varUpsilon="0107
\mathchardef\varPhi="0108
\mathchardef\varPsi="0109
\mathchardef\varOmega="010A
\def\wlog#1{\immediate\write-1{#1}}
\catcode`\@=12  
\def\=def{\; \mathop{=}_{\text{\rm def}} \;}
\def\rd{\partial}
\def\fhat{{\hat{f}}}
\def\that{{\hat{t}}}
\def\uhat{{\hat{u}}}
\def\vhat{{\hat{v}}}
\def\Nhat{{\hat{N}}}
\def\Shat{{\hat{S}}}
\def\calB{{\cal B}}
\def\calL{{\cal L}}
\def\calM{{\cal M}}
\def\calP{{\cal P}}
\def\calQ{{\cal Q}}
\def\calBhat{{\hat{\calB}}}
\def\calLhat{{\hat{\calL}}}
\def\calMhat{{\hat{\calM}}}
\def\res{\;\mathop{\text{res}}}
%
\hsize=15.5truecm
\vsize=23truecm
\sequentialequations
\doublespace
\TagsOnRight
\overfullrule=0pt
\sectionstyle={\Number}
\pubnum={KUCP-0061/93}
\date={May, 1993}
\titlepage
\title{\fourteencp
  Integrable hierarchy underlying topological Landau-Ginzburg models
  of D-type}
\author{Kanehisa Takasaki}
\address{
  Department of Fundamental Sciences\break
  Faculty of Integrated Human Studies, Kyoto University\break
  Yoshida-Nihonmatsu-cho, Sakyo-ku, Kyoto 606, Japan\break
  E-mail: takasaki @ jpnyitp (Bitnet)\break}

\abstract
\noindent
A universal integrable hierarchy underlying topological
Landau-Ginzburg models of D-type is presented.  Like
the dispersionless Toda hierarchy, the new hierarchy
has two distinct (``positive" and ``negative") set of
flows.  Special solutions corresponding to topological
Landau-Ginzburg models of D-type are characterized by
a Riemann-Hilbert problem, which can be converted into a
generalized hodograph transformation.  This construction
gives an embedding of the finite dimensional small phase
space of these models into the full space of flows of this
hierarchy.  One of flat coordinates in the small phase
space turns out to be identical to the first ``negative"
time variable of the hierarchy, whereas the others belong
to the ``positive" flows.

\endpage

\section{Introduction}

\noindent
Recent progress in Landau-Ginzburg models of topological strings
[\REF\DVV{
Dijkgraaf, R., Verlinde, E., and Verlinde, H.,
Topological strings in $d < 1$,
Nucl. Phys. B352 (1991), 59-86.
}\DVV]
has revealed an unexpected link with theories of singularities
and integrable hierarchies.  In the absence of gravitational
couplings (i.e., in the so called ``small phase space"),
these models of topological strings are reduced to a class of
topological conformal field theories.  Sophisticated
mathematical concepts in singularity theory, such as
``flat coordinates," ``higher residue pairings" and
``Gauss-Manin systems "
[\REF\SaitoNoumi{
K. Saito,
On the period of primitive integrals,
Publ. RIMS, Kyoto Univ., 19 (1983), 1231.\nextline
M. Noumi,
Expansion of the solutions of a Gauss-Manin system
at a point of infinity,
Tokyo J. Math. 7 (1984), 1-60.
}\SaitoNoumi]
are now recognized as very useful tools for studying this
class of topological conformal field theories
[\REF\BlokVarchenko{
Blok, B., and Varchenko, A.,
Topological conformal field theories and the flat coordinates,
Int. J. Mod. Phys. A7 (1992), 1467-1490.
}\BlokVarchenko].
These tools are also recently applied to an explicit
construction of ``gravitational descendents" in topological
strings
[\REF\LosevEguchietal{
Losev, A., Descendents constructed from matter field
in topological Landau-Ginzburg theories to topological gravity,
ITEP preprint (November, 1992). \nextline
Eguchi, T., Kanno, H., Yamada, Y., and Yang, S.-K.,
Topological strings, flat coordinates and gravitational
descendents,
University of Tokyo preprint UT-630 (February, 1993).
}\LosevEguchietal].
If gravitational couplings are turned on, flows in the space
of gravitational couplings obey an integrable hierarchy of
``Whitham type" or ``hydrodynamic type"
[\REF\KricheverDubrovin{
Krichever, I.M.,
The dispersionless Lax equations and topological minimal models,
Commun. Math. Phys. 143 (1991), 415-426. \nextline
Dubrovin, B.A.,
Hamiltonian formalism of Whitham-type hierarchies
and topological Landau-Ginsburg models,
Commun. Math. Phys. 145 (1992), 195-207.
}\KricheverDubrovin].
This is a place where the theory of integrable hierarchies
plays a key role.

Relation to integrable hierarchies takes a particularly
simple form in topological strings of A-type and D-type [\DVV].
The simplest and most well understood is the case of models
of A-type.  These models are realized as special solutions
of the dispersionless generalized KdV hierarchies.  Thus the
dispersionless KP hierarchy
[\REF\dKP{
Lebedev, D., and Manin, Yu.,
Conservation laws and Lax representation
of Benny's long wave equations,
Phys.Lett. 74A (1979), 154-156.\nextline
Kodama, Y.,
A method for solving the dispersionless KP equation and
its exact solutions,
Phys. Lett. 129A (1988), 223-226.\nextline
Kodama, Y., and Gibbons, J.,
A method for solving the dispersionless KP hierarchy and
its exact solutions, II,
Phys. Lett. 135A (1989), 167-170.
}\dKP]
emerges as a universal hierarchy of these models.
The second simplest case consists of models of D-type.
The Landau-Ginzburg potential of D-type contains two
fields, but one of them can easily be eliminated by
simple Gaussian path integral.  Apart from an extra term,
the reduced potential in the small phase space is
essentially a special case (or an ``orbifold") of A-type
models. Flows in the space of gravitational couplings,
however, remain to be studied. Naturally, no universal
integrable hierarchy has been identified.

We present in this paper a new integrable hierarchy that
underlies these topological Landau-Ginzburg models of D-type.
Our construction implements, from the very beginning,
an infinite number of integrable flows besides those in
the small phase space. These flows should be identified
with flows of gravitational couplings.  Actually, the
universal hierarchy is comprised of two distinct sets
of flows (so to speak, ``positive" and ``negative" flows)
just like the dispersionless Toda hierarchy
[\REF\TTdToda{
Takasaki, K., and Takebe, T.,
SDiff(2) Toda equation -- hierarchy, tau function and symmetries,
Lett. Math. Phys. 23 (1991), 205-214.
}\TTdToda].
Remarkably, an exceptional coordinate ``$t_*$" [\DVV] of the
small phase space can be identified with the first negative
flow; this clearly explains why it is exceptional.

Even apart from the relation to topological Landau-Ginzburg
medels, our new hierarchy possesses in itself a number of
interesting aspects as we now show below. Let us start from
exhibiting general properties, then turn to topological
Landau-Ginzburg models.

\section{Lax formalism of new hierarchy}

\noindent
The new hierarchy resemble both the dispersionless Toda
hierarchy [\TTdToda] and the dispersionless BKP hierarchy
[\REF\dBKP{
Takasaki, K.,
Quasi-classical limit of BKP hierarchy
and W-infinity symmetries,
Kyoto University preprint KUCP-0058/93  (January, 1993).
}\dBKP],
lying thus in between. (Actually, the KP hierarchy has a
C-type version, CKP, besides the B-type version
[\REF\DJKMIV{
Date, E., Jimbo, M., Kashiwara, M., and Miwa, T.,
Transformation theory for soliton equations IV,
J. Phys. Soc. Japan 50 (1982), 3813-3818.
}\DJKMIV].
They turn out to have substantially the same dispersionless
limit.) Its Lax representation, like the dispersionless Toda
hierarchy, consists of four sets of Lax equations of the form
$$
\align
    \dfrac{\rd \calL}{\rd t_{2n+1}} = \{ \calB_{2n+1}, \calL \},
    \quad&
    \dfrac{\rd \calL}{\rd \that_{2n+1}}
                    = \{ \calBhat_{2n+1}, \calL \},
                                                                 \\
    \dfrac{\rd \calLhat}{\rd t_{2n+1}} = \{ \calB_{2n+1}, \calLhat \},
    \quad&
    \dfrac{\rd \calLhat}{\rd \that_{2n+1}}
                    = \{ \calBhat_{2n+1}, \calLhat \},
    \quad n = 0,1, \ldots,
                                                     \tag\eq     \\
\endalign
$$
describing two sets of flows in two sets of time variables
$t = (t_1,t_3,\ldots)$ and $\that = (\that_1,\that_3,\ldots)$.
The classical counterparts of $L$-operators
$$
\align
    \calL =& k + \sum_{n=1}^\infty u_{2n} k^{-2n+1},    \\
    \calLhat =& \sum_{n=0}^\infty \uhat_{2n} k^{2n+1},
    \quad  \uhat_0 \not= 0,                          \tag\eq
                                                        \\
\endalign
$$
are Laurent series in another variable $k$. The classical
counterparts of $B$-operators are given by
$$
    \calB_{2n+1} = \left( \calL^{2n+1} \right)_{\ge 0}, \quad
    \calBhat_{2n+1} = \left( \calLhat^{-2n-1} \right)_{\le -1},
                                                     \tag\eq
$$
where
$$
\align
  & (\quad)_{\ge 0}:  \text{projection onto } k^0,k^1,\ldots,   \\
  & (\quad)_{\le -1}: \text{projection onto } k^{-1},k^{-2},\ldots.
                                                        \tag\eq  \\
\endalign
$$
Unlike the dispersionless Toda hierarchy, however, $\calL$ and
$\calLhat$ contains only odd powers of $k$. This is a reason that
only ``odd" time variables are permitted. Furthermore, the
Poisson bracket, too, is different and given by
$$
    \{ A, B \} = \frac{\rd A}{\rd k} \frac{\rd B}{\rd x}
                -\frac{\rd A}{\rd x} \frac{\rd B}{\rd k}, \quad
    x = t_1.                                              \tag\eq
$$
(The Poisson bracket in the dispersionless Toda hierarchy is
defined for a rescaled lattice coordinate $s$ and its canonical
momentum $p$) These characteristics are rather reminiscent of
the dispersionless BKP (or CKP) hierarchy. In fact, the first
set of Lax equations, including $\calL$ and $t$ only, is nothing
but the dispersionless BKP hierarchy. The new hierarchy is thus
a straightforward extension of the dispersionless BKP hierarchy
by the ``negative" flows $\that$.

As in other dispersionless integrable hierarchies, one can
now introduce another set of Laurent series
$$
\align
   \calM =&  \sum_{n=0}^\infty (2n+1) t_{2n+1} \calL^{2n}
           + \sum_{n=0}^\infty v_{2n+2} \calL^{-2n-2},      \\
   \calMhat
   =& - \sum_{n=0}^\infty (2n+1) \that_{2n+1} \calLhat^{-2n-2}
      + \sum_{n=0}^\infty \vhat_{2n+2} \calLhat^{2n}
                                  \tag\eqname\MMhatInLLhat  \\
\endalign
$$
that satisfy the Lax equations
$$
\align
    \dfrac{\rd \calM}{\rd t_{2n+1}} = \{ \calB_{2n+1}, \calM \},
    \quad &
    \dfrac{\rd \calM}{\rd \that_{2n+1}}
                    = \{ \calBhat_{2n+1}, \calM \},
                                                                 \\
    \dfrac{\rd \calMhat}{\rd t_{2n+1}} = \{ \calB_{2n+1}, \calMhat \},
    \quad &
    \dfrac{\rd \calMhat}{\rd \that_{2n+1}}
                    = \{ \calBhat_{2n+1}, \calMhat \},
    \quad n = 0,1, \ldots
                                                     \tag\eq     \\
\endalign
$$
and the canonical Poisson relations
$$
    \{ \calL, \calM \} = \{ \calLhat, \calMhat \} =1.    \tag\eq
$$
Note that these $\calM$ and $\calMhat$ contain only even powers
of $\calL$ and $\calLhat$. (This definition of $\calM$ slightly
differs from our previous notations for the dispersionless BKP
hierarchy [\dBKP]; the above $\calM$ amounts to $\calM\calL^{-1}$
in the notation used therein.) As already pointed out in the case
of other dispersionless hierarchies and selfdual gravity
[\REF\Turku{
Takasaki, K.,
Area-preserving diffeomorphisms and nonlinear integrable systems,
in: {\it Topological and geometrical methods in field theory\/},
Turku, Finland, 1991, ed. J. Mickelsson and O. Pekonen
(World Scientific, Singapore, 1992).
}\Turku],
this kind of extended Lax formalism is a key to a  Riemann-Hilbert
problem method.

\section{Riemann-Hilbert problem}

\noindent
The extended Lax representation can be rewritten into the
2-form equation
$$
   d\calL \wedge d\calM = d\calLhat \wedge d\calMhat = \omega,
                                           \tag\eqname\TwoFormEq
$$
where
$$
   \omega = \sum_{n=0}^\infty d\calB_{2n+1} \wedge dt_{2n+1}
    + \sum_{n=0}^\infty d\calBhat_{2n+1} \wedge d\that_{2n+1}.
                                                    \tag\eq
$$
Note that $\omega$ includes the symplectic 2-form in the
$(k,x)=(k,t_1)$ space:
$$
   d\calB_1 \wedge dt_1 = dk \wedge dx.             \tag\eq
$$

The above 2-form equation implies that $(\calL,\calM)$ and
$(\calLhat,\calMhat)$ are two sets of ``Darboux coordinates"
of $\omega$ and linked by a functional relation of the form
(``Riemann-Hilbert problem")
$$
   \calLhat = f(\calL,\calM), \quad
   \calMhat = g(\calL,\calM),
                                          \tag\eqname\RHGen
$$
where $f(\lambda,\mu)$ and $g(\lambda,\mu)$ are subject to
the symplectic condition
$$
   \dfrac{\rd \bigl( f(\lambda,\mu), g(\lambda,\mu) \bigr) }
         {\rd \bigl( \lambda, \mu \bigr) }
   = 1,                                             \tag\eq
$$
i.e., gives a symplectic (or area-preserving) diffeomorphism.
Furthermore, in view of the parity of $\calL$, $\calLhat$ (odd)
and $\calM$, $\calMhat$ (even) under $k \to -k$,
one has to impose the extra condition
$$
    f(-\lambda,\mu) = - f(\lambda,\mu), \quad
    g(-\lambda,\mu) = g(\lambda,\mu).               \tag\eq
$$
The last condition, conversely, ensures that solutions of
the above Riemann-Hilbert problem indeed give solutions of
the hierarchy in question (as far as only ``odd" time variables
are retained).

\section{$S$-functions and free energy}

\noindent
Our goal here is to show the existence of a potential $F$
(``free energy") along with two auxiliary functions
$S$ and $\Shat$. The construction is parallel to those of
the dispersionless Toda hierarchy.  A key role is played by
the residue formulas
$$
\align
    \dfrac{\rd v_{2m+2}}{\rd t_{2n+1}}
    = \res_k \left( \calL^{2m+1} \rd_k \calB_{2n+1} \right),
     \quad&
    \dfrac{\rd v_{2m+2}}{\rd \that_{2n+1}}
    = \res_k \left( \calL^{2m+1} \rd_k \calBhat_{2n+1} \right),
                                                               \\
    \dfrac{\rd \vhat_{2m+2}}{\rd t_{2n+1}}
    = \res_k \left( \calLhat^{-2m-1} \rd_k \calB_{2n+1} \right),
      \quad&
    \dfrac{\rd \vhat_{2m+2}}{\rd \that_{2n+1}}
    = \res_k \left( \calLhat^{-2m-1} \rd_k \calBhat_{2n+1} \right),
                                   \tag\eqname\ResidueFormula  \\
\endalign
$$
where ``$\res_k$" means the formal residue in $k$:
$$
    \res_k \sum a_n k^n = a_{-1}.                 \tag\eq
$$
Technical details for deriving these and subsequent formulas
are almost the same as the case of the dispersionless KP hierarchy
[\REF\TTdKP{
Takasaki, K., and Takebe, T.,
SDiff(2) KP hierarchy,
in: {\it Infinite Analysis\/}, RIMS Research Project 1991,
Int. J. Mod. Phys. A7, Suppl. 1
(World Scientific, Singapore, 1992).
}\TTdKP].

As a consequence of (\ResidueFormula) and a few basic
properties of the formal residue operator, one can show
the relations
$$
\align
  & \dfrac{\rd v_{2m+2}}{\rd t_{2n+1}}
             = \dfrac{\rd v_{2n+2}}{\rd t_{2m+1}},
    \quad
    \dfrac{\rd \vhat_{2m+2}}{\rd \that_{2n+1}}
             = \dfrac{\rd \vhat_{2n+2}}{\rd \that_{2m+1}},
                                                           \\
  & \dfrac{\rd v_{2m+2}}{\rd \that_{2n+1}}
             = - \dfrac{\rd \vhat_{2n+2}}{\rd t_{2m+1}}.
                                                 \tag\eq   \\
\endalign
$$
This implies the existence of a potential $F$ that reproduces
$v_{2n}$ and $\vhat_{2n}$ as:
$$
    \dfrac{\rd F}{\rd t_{2n+1}} = v_{2n+2},            \quad
    \dfrac{\rd F}{\rd \that_{2n+1}} = - \vhat_{2n+2}.  \tag\eq
$$

Another consequence of (\ResidueFormula) is an explicit expression
of of $\calB_{2n+1}$ and $\calBhat_{2n+1}$ as Laurent series of
$\calL$ and $\calLhat$.  To see this, let us consider the formal
residue operators $\res_\calL$ and $\res_\calLhat$ with respect
to $\calL$ and $\calLhat$. They are connected with $\res_k$ as:
$$
\align
    & \res_\calL \calL^n = \res_k \calL^n \rd_k \calL
                         = \delta_{n,-1},                      \\
    & \res_\calLhat \calLhat^n = \res_k \calLhat^n \rd_k \calLhat
                         = \delta_{n,-1}.       \tag\eq        \\
\endalign
$$
By use of these relations, one can derive from (\ResidueFormula)
the following formulas.
$$
\align
  \calB_{2n+1}
  =& \calL^{2n+1}
     - \sum_{m=0}^\infty \frac{1}{2m+1}
        \dfrac{\rd v_{2m+2}}{\rd t_{2n+1}} \calL^{-2m-1}         \\
  =& \sum_{m=0}^\infty \frac{1}{2m+1}
        \dfrac{\rd \vhat_{2m+2}}{\rd t_{2n+1}} \calLhat^{2m+1},
                                                                 \\
  \calBhat_{2n+1}
  =& -\sum_{m=0}^\infty \frac{1}{2m+1}
        \dfrac{\rd v_{2m+2}}{\rd \that_{2n+1}} \calL^{-2m-1}     \\
  =& \calLhat^{-2n-1}
     + \sum_{m=0}^\infty \frac{1}{2m+1}
       \dfrac{\rd \vhat_{2m+2}}{\rd \that_{2n+1}} \calLhat^{2m+1}.
                                    \tag\eqname\BBhatInLLhat     \\
\endalign
$$
In particular, one obtains two expressions of $\calB_1 = k$:
$$
   \calB_1
   = \calL + \sum_{n=1}^\infty f_{2n} \calL^{-2n+1}
   = \sum_{n=0}^\infty \fhat_{2n} \calLhat^{2n+1}    \tag\eq
$$
where
$$
\align
   f_{2n} =& - \frac{1}{2n-1}
                 \dfrac{\rd^2 F}{\rd t_1 \rd t_{2n-1}},
                                                                     \\
   \fhat_{2n} =& - \frac{1}{2n+1}
                     \dfrac{\rd^2 F}{\rd t_1 \rd \that_{2n+1}}.
                                                     \tag\eq         \\
\endalign
$$
Actually, these expressions of $B_1 =k$ give an inversion of
the maps $k \to \calL$ and $k \to \calLhat$ respectively.
Solving them with respect to $\calL$ and $\calLhat$ once again,
one can express $u_{2n}$ and $\uhat_{2n}$ in terms of derivatives
of $F$. For instance,
$$
    u_2 = - f_2 = \dfrac{\rd^2 F}{\rd t_1^2}, \quad
    \uhat_0^{-1} = \fhat_0 = - \frac{\rd^2 F}{\rd t_1 \rd \that_1}.
                                                     \tag\eq
$$
The above definition of $f_{2n}$ and $\fhat_{2n}$ can be rewritten
$$
\align
   f_{2n} =& - \frac{1}{2n-1} \res_k \calL^{2n-1},
                                                                     \\
   \fhat_{2n} =& \frac{1}{2n+1} \res_k \calLhat^{-2n-1}.
                                                     \tag\eq         \\
\endalign
$$
We shall see in the next section that these quantities are
directly related to flat coordinates of Landau-Ginzburg models.

Finally, let us define the $S$-functions
$$
\align
  S =& \sum_{n=0}^\infty t_{2n+1} \calL^{2n+1}
     - \sum_{n=0}^\infty \dfrac{v_{2n+2}}{2n+1} \calL^{-2n-1},
                                                                  \\
  \Shat =& \sum_{n=0}^\infty \that_{2n+1} \calLhat^{-2n-1}
    + \sum_{n=0}^\infty \dfrac{\vhat_{2n+2}}{2n+1} \calLhat^{2n+1}.
                                                     \tag\eq      \\
\endalign
$$
As a consequence of (\MMhatInLLhat) and (\BBhatInLLhat),
they satisfy the 1-form equations
$$
\align
    dS =& \calM d \calL
        + \sum_{n=0}^\infty \calB_{2n+1} dt_{2n+1}
        + \sum_{n=0}^\infty \calBhat_{2n+1} d\that_{2n+1},
                                                                  \\
    d\Shat =& \calMhat d \calLhat
        + \sum_{n=0}^\infty \calB_{2n+1} dt_{2n+1}
        + \sum_{n=0}^\infty \calBhat_{2n+1} d\that_{2n+1}.
                                                                  \\
\endalign
$$
Exterior differentiation of these equations reproduces 2-form
equation (\TwoFormEq).

\section{Topological Landau-Ginzburg models}

\noindent
The $\text{D}_N$ Landau-Ginzburg potential is a polynomial of
two fields $X$ and $Y$:
$$
   W(X,Y) = \frac{X^{N-1}}{2N-2} + \frac{1}{2} X Y^2
           + \sum_{n=1}^{N-1} g_{2n} X^{N-n-1} + t_{*} Y,
                                                       \tag\eq
$$
where $g_2$, $\ldots$, $g_{2N-2}$ depend on $N-1$ deformation
variables $(t_1,t_3,\ldots,t_{2N-3})$; $t_{*}$ is the last
deformation variable. (Our notation slightly differs from
physicists' convention. Our $t_1$, $\ldots$, $t_{2N-3}$
correspond to physicists' $t_0$, $\ldots$, $t_{2N-4}$,
up to suitable rescaling.) As pointed out by Dijkgraaf et al.
[\DVV], the field $Y$ can be eliminated by Gaussian path
integral or, equivalently, by solving the relation
$\rd_Y W(X,Y) = 0$.  Upon substituting
$$
    X = Z^2,                                           \tag\eq
$$
the reduced potential is given by
$$
    W_{\text{red}}(Z) = \frac{Z^{2N-2}}{2N-2}
       + \sum_{n=1}^{N-1} g_{2n} Z^{2N-2n-2}
       + g_{2N}Z^{-2},                                 \tag\eq
$$
where
$$
    g_{2N} = - \frac{1}{2} t_{*}^2.     \tag\eqname\GAndTstar
$$
The reduced potential consists of two parts,
$$
    W_{\text{red}}(Z) = W_0(Z) + g_{2N}Z^{-2},         \tag\eq
$$
the first part $W_0(Z)$ being interpreted as a subfamily of
deformations of the $\text{A}_{2N-3}$ Landau-Ginzburg potential.
The first $N-1$ deformation variables $(t_1,\ldots,t_{2N-1})$
are identical, up to rescaling by constants, to the corresponding
$\text{A}_{2N-3}$ flat coordinates. The status of $t_{*}$ remains
rather obscure in the work of Dijkgraaf et al.

We now show how to interpret the reduced Landau-Ginzburg
model $W_{\text{red}}(Z)$ as a special solution of our
hierarchy.  A very similar interpretation is recently
presented by Eguchi et al.
[\REF\Eguchietal{
Eguchi, T., Yamada, Y., and Yang, S.-K.,
Topological field theories and the period integrals,
University of Tokyo preprint UT-641 (April, 1993).
}\Eguchietal],
however their treatment is limited to the small phase space.
Our construction of solution is a generalization of our
previous treatment of A-type models [\TTdKP],
and can cover all $(t,\that)$ flows simultaneously.
A Riemann-Hilbert problem lies in the heart of this
construction.

The Riemann-Hilbert problem in this case takes the following form.
$$
    \dfrac{\calL^{2N-2}}{2N-2} = \dfrac{\calLhat^{-2}}{-2}
        \quad ( \=def \calP ),
    \qquad
    \calM \calL^{-2N+3} = \calMhat \calLhat^3
        \quad ( \=def \calQ ).
                                          \tag\eqname\FunctEq
$$
This Riemann-Hilbert problem does not take the standard form of
(\RHGen), but the essence is the same. Note that the maps
$(\calL,\calM)$ $\to$ $(\calP,\calQ)$ and
$(\calLhat,\calMhat)$ $\to$ $(\calP,\calQ)$ are both symplectic.
Therefore, if one can solve the above Riemann-Hilbert problem,
the four functions $\calL$, $\calM$, $\calLhat$, $\calMhat$
satisfy the relation
$$
    d\calL \wedge d\calM = d\calP \wedge d\calQ
    = d\calLhat \wedge d\calMhat,                   \tag\eq
$$
which implies that they give a solution of the hierarchy.

The above Riemann-Hilbert problem, like the case of A-type models,
can be reduced to a set of equations whose solvability is ensured
by the implicit function theorem. The first equation of (\FunctEq)
simply means that $\calP$ is an even Laurent polynomial of $k$
of the form
$$
    \calP = \dfrac{k^{2N-2}}{2N-2}
            + \sum_{n=1}^{N} g_{2n} k^{2N-2n-2}.      \tag\eq
$$
The coefficients of $\calP$, $\calL$ and $\calLhat$ obey
algebraic relations of the form
$$
\align
  & g_2 = u_2, \quad
    g_4 = u_4 + \frac{2N-3}{2} u_2^2, \ldots,              \\
                                                           \\
  & g_{2N} = -\frac{1}{2}\uhat_0^{-2},  \quad
    g_{2N-2} = \uhat_0^{-3} \uhat_2, \ldots,
                                        \tag\eqname\GAndUUhat
                                                           \\
\endalign
$$
which can uniquely be solved for $u_{2n}$ and $\uhat_{2n}$. The
second equation of (\FunctEq) requires technically more involved
calculations. We first separate it into the $(\quad)_{\le -1}$-part
and the $(\quad)_{\ge 0}$-part. This results in the following
equations.
$$
\align
   &  \sum_{n=0}^{N-2} t_{2n+1} \calL^{2n-2N+3}
     + \sum_{n=1}^\infty v_{2n} \calL^{-2n-2N+3}               \\
  =& - \sum_{n=N-1}^\infty (2n+1)
            t_{2n+1} (\calL^{2n-2N+3})_{\le -1}
     - \sum_{n=1}^\infty
            (2n+1) \that_{2n+1} ( \calLhat^{-2n+1})_{\le -1},
                                      \tag{\eqname\FunctEqQ a} \\
   & - \that_1 \calLhat
     + \sum_{n=1}^\infty \vhat_{2n} \calLhat^{2n+1}            \\
  =& - \sum_{n=N-1}^\infty (2n+1)
            t_{2n+1} (\calL^{2n-2N+3})_{\ge 0}
     - \sum_{n=1}^\infty
            (2n+1) \that_{2n+1} ( \calLhat^{-2n+1})_{\ge 0}.
                                       \tag{\FunctEqQ b}       \\
\endalign
$$
By means of formal residue calculus, we can rewrite these equations
into equations of Laurent coefficients. Eq. ({\FunctEqQ}a) can thus
be decomposed into
$$
\align
  (2n+1) t_{2n+1}
  =& - \sum_{m=N-1}^\infty (2m+1) t_{2m+1} \res_k \left[
      (\calL^{2m-2N+3})_{\le -1} \calL^{-2n+2N-4} \rd_k \calL \right] \\
   & - \sum_{m=1}^\infty (2m+1) \that_{2m+1} \res_k \left[
      (\calLhat^{-2m+1})_{\le -1} \calL^{-2n+2N-4} \rd_k \calL \right]
                                                 \tag\eqname\HodoEqT
                                                                      \\
\endalign
$$
for $n=0,1,\ldots,N-2$, and
$$
\align
  v_{2n}
  =& - \sum_{m=N-1}^\infty (2m+1) t_{2m+1} \res_k \left[
      (\calL^{2m-2N+3})_{\le -1} \calL^{2n+2N-4} \rd_k \calL \right] \\
   & - \sum_{m=1}^\infty (2m+1) \that_{2m+1} \res_k \left[
      (\calLhat^{-2m+1})_{\le -1} \calL^{2n+2N-4} \rd_k \calL \right]
                                                 \tag\eqname\HodoEqV
                                                                     \\
\endalign
$$
for $n=1,2,\ldots$.  Eq. ({\FunctEqQ}b) can, similarly, be decomposed
into
$$
\align
  \that_1
  =& - \sum_{m=N-1}^\infty (2m+1) t_{2m+1} \res_k \left[
     (\calL^{2m-2N+3})_{\ge 0} \calLhat^{-2} \rd_k \calLhat \right] \\
   & - \sum_{m=1}^\infty (2m+1) \that_{2m+1} \res_k \left[
     (\calLhat^{-2m+1})_{\ge 0} \calLhat^{-2} \rd_k \calLhat \right]
                                              \tag\eqname\HodoEqThat
                                                                     \\
\endalign
$$
and
$$
\align
  \vhat_{2n}
  =& \sum_{m=N-1}^\infty (2m+1) t_{2m+1} \res_k \left[
  (\calL^{2m-2N+3})_{\ge 0} \calLhat^{-2n-2} \rd_k \calLhat \right] \\
   & +\sum_{m=1}^\infty (2m+1) \that_{2m+1} \res_k \left[
     (\calLhat^{-2m+1})_{\ge 0} \calLhat^{-2n-2} \rd_k \calLhat \right]
                                              \tag\eqname\HodoEqVhat
                                                                     \\
\endalign
$$
for $n=1,2,\ldots$.
These equations give a D-type version of the generalized hodograph
transformation of A-type models [\TTdKP]. The first half, (\HodoEqT)
and (\HodoEqThat), of these equations are to determine $g_2$, $\ldots$,
$g_{2N}$ (hence $\calL$ and $\calLhat$, too) as implicit functions of
$t$ and $\that$. The implicit function theorem indeed ensures the
existence of a unique solution in, for instance, a neighborhood of
the submanifold
$$
\align
   & t_{2N-1} = - \frac{1}{2N-1}, \quad
     t_{2N+1} = t_{2N+3} = \cdots = 0,                              \\
   & \that_3 = \that_5 = \cdots = 0.          \tag\eqname\SmallSpace
                                                                    \\
\endalign
$$
The rest of Eqs. (\HodoEqV) and (\HodoEqVhat), give a considerably
complicated, but explicit expression of $v_{2n}$ and $\vhat_{2n}$.

It is exactly on this submanifold, (\SmallSpace), that the reduced
Landau-Ginzburg potential $W_{\text{red}}(Z)$ is reproduced upon
identifying $Z = k$ and $W_{\text{red}}(Z) = \calP$.  In other words,
this submanifold is nothing but the small phase space of the
$\text{D}_N$ model. To see this, note that Eqs. (\HodoEqT) and
(\HodoEqThat) are simplified on this submanifold as:
$$
\align
  &  (2n+1) t_{2n+1} = - f_{2N-2n-2} \quad (n=0,\ldots,N-2),  \\
  &  \that_1 = \fhat_0 = \uhat_0^{-1}.
                                                   \tag\eq    \\
\endalign
$$
The first relation tells that the $t$ variables coincide, up to
rescaling constants, with flat coordinates of the $\text{A}_{2N-3}$
model restricted to the subspace $t_{2n} = 0$, thus reproducing
the result of Dijkgraaf et al. [\DVV]. The second relation,
meanwhile, shows that
$$
    t_{*} = \that_1.                  \tag\eqname\TstarAndThat
$$
[Compare the two expressions of $g_{2N}$ in (\GAndTstar) and
(\GAndUUhat). In view of the above relation between $\that_1$
and $\uhat_0$, one can readily derive (\TstarAndThat).]  Thus,
the somewhat distinct deformation variable $t_{*}$ turns out
to be identical to the first ``negative" time of our hierarchy.

Let us finally note that the above construction can be extended
straightforward to a more general case (with two discrete
parameters $N$ and $\Nhat$) of the form
$$
    \dfrac{\calL^{2N-2}}{2N-2}
    = \dfrac{\calLhat^{-2\Nhat+2}}{-2\Nhat+2}, \quad
    \calM \calL^{-2N+3}
    = \calMhat \calLhat^{2\Nhat-1}.
                                      \tag\eqname\FunctEqNNhat
$$
In view of the general classification program of topological
conformal field theories by  Krichever and Dubrovin
[\KricheverDubrovin], these solutions, too, deserve to be
studied from a physical point of view.

\section{Conclusion}

\noindent
Our results are divided into two parts.  In the first part, the
new hierarchy is introduced and general properties are specified.
The hierarchy itself, in one hand, resembles the dispersionless
Toda hierarchy in the sense that it consists of two distinct
sets of flows.  The ``positive" flows, on the other hand, are
substantially the same as the dispersionless BKP (or CKP)
hierarchy.  In the second part, topological Landau-Ginzburg
models of D-types are identified with special solutions of
this hierarchy. These solutions are characterized by a
Riemann-Hilbert problem, which can be converted into a
generalized hodograph transformation.

Several new aspects of D-type models are revealed in these results.
For instance, one can clarify the status of the somewhat distinct
deformation variable $t_{*}$ in these models. In fact, $t_{*}$
turns out to be identical to the first ``negative" time variable
of the hierarchy, whereas the other variables $t_1$, $\ldots$,
$t_{2N-3}$ belong to the ``positive" flows.  The finite
dimensional small phase space with these ``flat coordinates"
are now embedded into the full space of flows of the hierarchy.
The time variables other than $t_1$, $\ldots$, $t_{2N-3}$ and
$\that_1$ should be interpreted as gravitational couplings.
Furthermore, although lacking physical interpretation,
more general solutions with two discrete parameters are shown
to be constructed in the same manner.

We add a few comments.

1. As already pointed out by Dijkgraaf et al. [\DVV], topological
Landau-Ginzburg models of D-type are related to Drinfeld-Sokolov
hierarchies of D-type. The hierarchy of $(D^{(1)}_N, c_0)$ type
in their table, indeed, turns out to reproduce the $t$ flows of
$W_{\text{red}}(Z)$ (or of $\calP$) in dispersionless (or
quasi-classical) limit.  Deriving the $\that$ flows remains to
be solved. This issue seems to require a suitable extension of
the Drinfeld-Sokolov theory.

2. An infinite dimensional Lie algebra called $D_\infty$ and
an associated integrable hierarchy are presented by Jimbo and Miwa
[\REF\JimboMiwa{
Jimbo, M., and Miwa, T.,
Solitons and infinite dimensional Lie algebras,
Publ. RIMS, Kyoto Univ., 19 (1983), 943-1001.
}\JimboMiwa].
This is a D-type version of the KP hierarchy. The $D^{(1)}_N$
algebras are included in $D_\infty$ as subalgebras, and
associated reductions are shown (at least at the level of
Hirota bilinear equations) to agree with the Drinfeld-Sokolov
$(D^{(1)}_N,c_0)$ hierarchies. Furthermore, remarkably, the
hierarchy consists of two sets of flows like the $(t,\that)$
flows of our hierarchy.  Thus, the hierarchy of Jimbo and Miwa
will presumably be a ``dispersionful" counterpart of our
hierarchy.

The author would like to thank Takashi Takebe for many useful
comments. He is also grateful to Dr. Yasuhiko Yamada for
hospitality at a KEK seminar on topological field theories
and period integrals. This work is partly supported by the
Grant-in-Aid for Scientific Research, the Ministry of Education,
Science and Culture, Japan.

\refout
\bye